\documentstyle[12pt]{article}
\begin{document}
\parskip=6pt
\baselineskip=20pt
{\raggedleft{AS-ITP-95-16\\}}
{\raggedleft{April 29, 1995\\}}
\bigskip
\centerline{\Large\bf Three-Dimensional Vertex Model Related}
\centerline{\Large\bf BCC Model in Statistical Mechanics}
\bigskip
\centerline{\large\bf Zhan-Ning Hu \footnote{\bf email address:
huzn@itp.ac.cn}}
\smallskip
\centerline{CCAST(WORLD LABORATORY) P.O.BOX 8730, BEIJING, 100080}
\smallskip
\centerline{and}
\centerline{INSTITUTE OF THEORETICAL PHYSICS, ACADEMIA SINICA,}
\centerline{ P. O. BOX 2735 BEIJING 100080, CHINA \footnote{\bf
mail address}}
\vspace{2ex}
\vspace{2ex}
\bigskip
\begin{center}
\begin{minipage}{5in}
\centerline{\large\bf 	Abstract}
\vspace{1ex}
In this paper, a three-dimensional vertex model is obtained. It is
a duality of the three-dimensional integrable lattice model with $N$
 states proposed by Boos, Mangazeev, Sergeev and Stroganov. The Boltzmann
weight of the model is dependent on four spin variables, which are the
 linear combinations of the spins on the corner sites of the cube, and
 obeys the modified vertex type tetrahedron equation.  This vertex model
 can be regard as a deformation of the one related the three-dimensional
 Baxter-Bazhanov model. The constrained conditions of the spectrums  are
discussed in detail and the symmetry properties of weight functions of
the vertex model are presented.

\bigskip

\end{minipage}
\end{center}
\newpage
\section{\bf Introduction}

There is a large class of integrable lattice models in statistical
mechanics.  Most of them are two-dimensional, such as Ising model
 \cite{44}, six-vertex model \cite{lieb}, eight-vertex model
 \cite{book}, chiral Potts model \cite{prprint} $et~al$,  where the
Yang-Baxter relation (or the star-triangle relation) plays an important
 role. It ensures the commutativity of the transfer-matrices, then
 permits us to calculate the partition function per site in the
thermodynamical limit. Recently much attention has been payed on
 the three-dimensional integrable lattice model.  Bazhanov and Baxter
generalized the Zamolodchikov's two states model \cite{Zam} to the
 arbitrary states \cite{BB1} and Kashaev $ et~al$ showed that the
tetrahedron equation holds in this model by introducing the star-square
 relation \cite{Kas2,huremark,humodlett,hulettA2}. The symmetry properties
 of the weight functions have been discussed in ref \cite{Kas1,hu1,BB2}.
 Boos and Mangazeev $et~al$ obtained another kind of the lattice model
 in three-dimensions.

One of the other approaches of the three-dimensional integrable models
 is of  the vertex typed models. Hietarinta discussed the tetrahedron
equation by considering the scattering of straight strings \cite{Hir},
 similarly as in ref \cite{Zam}, where  the three labeling schemes exist.
 The duality between them is discussed in detail. Then the main problems
 are to solved these tetrahedron equations and to build the integrable
 models. In fact, Korepanov \cite{Kor} given the solutions of the
vertex type tetrahedron equation for the two states in 1993, which
is different from the ones in \cite{Hir}. The vertex type tetrahedron
 equations are discussed also in refs. \cite{Viallet,Hielett,Horibe}.
 Another kind of the vertex model in three dimensions is proposed in
ref. \cite{newsolution}. Now we know that the Boltzmann weight of this
 model can be obtained from that of the Baxter-Bazhanov model
 \cite{humiwa}. And the three-dimensional vertex model as the
 correspondence of Baxter-Bazhanov model has been built in ref.
 \cite{hujsp}. In this paper, we obtained the three-dimensional
 vertex model corresponding to the lattice integrable model proposed
 in ref. \cite{boose,mangazeev}. It can be regard as a deformation
 of the three-dimensional vertex model in \cite{hujsp}. The connections
 of the  restricted conditions of the spectrums  between these models
 are discussed in detail and the symmetry properties are presented
 also. The weight function of the model depends on the $four$ spins
 variables and satisfies the modified vertex type tetrahedron equation.

The organization of this paper is as follows. In section 2, some
 notations and the necessary definitions are given and the weight
 functions of the lattice integrable model are presented. They satisfy
 the modified tetrahedron equations. In section 3, we find that the
 weight function of the model is dependent on $four$ spins variables
 which are the linear combinations of the spins located on the corner
 sites of the cubes.  We formulate the weight functions as the vertex
forms. They satisfy the modified vertex type tetrahedron equations.
 In section 4 we compare the weight functions between this model and
 the three-dimensional vertex model related the Baxter-Bazhanov model.
  The former can be regard as a deformation of the later. Then the
connections of the restricted conditions for the two kinds of vertex
 models are discussed in detail. In section 5, the symmetry properties
are presented, which are similar as in ref. \cite{hujsp}. As the short
 conclusions, some remarks are given finally.

\section{\bf The Body-Centered-Cube Model}

In this section, the brief describe of the lattice integrable model
 proposed in \cite{boose,mangazeev} is given and the necessary
definitions and some notations are presented. The weight functions
of this model have the Body-Centered-Cube form, similarly as the
 Baxter-Bazhanov model, and satisfy the modified tetrahedron equation
 which leads to the commuting family of two-layer transfer-matrices.
 The simple cubic lattice ${\cal L}$ is consisted of two types of
 elementary cubes alternating in checkerboard order in all directions
 and at each site of ${\cal L}$ place a spin with $Z_N$ values for
integer $N\geq 2$.

The weight function $W(a|efg|bcd|h)$ of Body-Center-Cube (BCC) model
 (see ref. \cite{boose}) depends on the eight surrounding spin
variables $a,b,\cdots,h$ and has the form
$$
W_P(a|efg|bcd|h) ~~~~~~~~~~~~~~~~~~~~~~~~~~~~~~~~~~~~~~~~~~~~~~~~~
{}~~~~~~~~~~~~~~~
$$$$
=\Bigg[\frac{w(x_{58}x_{67},x_8\mu,x_{13}x_{24}x_7x_8/x_3x_4|a+d,e+f)}
{w(x_{58}x_{67},x_5\nu,x_{13}x_{24}x_5x_6/x_1x_2|g+h,c+b)}\Bigg]^{1/2}~~~
$$$$
\times \Bigg[\frac{w(x_1x_8,x_1u,x_3x_5|e+h,d+c)}
{w(x_4x_6,x_2v,x_2x_7|a+b,f+g)}\Bigg]^{1/2} ~~~~~~~~~~~~~~~~~~
$$$$
\times \Bigg[\frac{w(x_2x_8,x_4\lambda,x_4x_5|e+g,a+c)}
{w(x_3x_6,x_3\xi,x_1x_7|b+d,f+h)}\Bigg]^{1/2}
\frac{\omega^{bf}\omega^{(ag+gb+bh)/2}}{\omega^{ag}
\omega^{(hd+de+ea)/2}}~
$$
\begin{equation} \label{W1}
{}~~~~~~~~~~~~~\times \Bigg\{\sum_{\sigma \in Z_N}
\frac{w(x_3,x_{13},x_1|d,h+\sigma)w(x_4,x_{24},x_2|a,g+\sigma)}
{w(x_8,x_{58},x_5|e,c+\sigma)
w(x_7/\omega,x_{67},x_6|f,b+\sigma)}\Bigg\}_0
\end{equation}
where  the subscript "0" after the curly brackets indicates that
 the  expression in the braces is divided by itself with the zero
 exterior spins. The function  $w(x,y,z|k,l)$ is defined as
\begin{equation}
w(x,y,z|k,l)=w(x,y,z|k-l)\Phi(l),~~w(x,y,z|l)=\prod^{l}_{j=1}
\frac{y}{z-x\omega^j},~~k,l\in Z_N
\end{equation}
with the notation
\begin{equation}
x^N+y^N=z^N,~~\Phi(l)=\omega^{l(l+N)/2},~~\omega^{1/2}=\exp(\pi i/N).
\end{equation}
The relation $x^N+y^N=z^N$ decides the variables
$x_{13},x_{24},x_{58},x_{67}$ and $ u,v,\xi,\lambda,\mu,\nu$
 up to some $\omega$ factors. The weight function satisfies the
 modified tetrahedron equation \cite{manstr}
$$
\sum_{d}
W(a_4|c_2c_1c_3|b_1b_3b_2|d)\stackrel{-}{W}'(c_1|b_2a_3b_1|c_4dc_6|b_4)
 ~~~~~~~~~~~~~~~~~~~~~~~~~~~~~~~~
$$$$
\times W''(b_1|dc_4c_3|a_2b_3b_4|c_5)
\stackrel{-}{W}'''(d|b_2b_4b_3|c_5c_2c_6|a_1)~~~~~~~~~~~~
{}~~~~~~~~~~~
$$$$
{}~~~~~~~~~~~~~~~~~=\sum_{d}
W'''(b_1|c_1c_4c_3|a_2a_4a_3|d)\stackrel{-}{W}''
(c_1|b_2a_3a_4|dc_2c_6|a_1)
$$
\begin{equation} \label{cube}
{}~~~~~~~~~~~~~~~~~~~~~~~~~~~\times W'(a_4|c_2dc_3|a_2b_3a_1|c_5)
\stackrel{-}{W}(d|a_1a_3a_2|c_4c_5c_6|b_4)
\end{equation}
where $W,W',W'',W'''$ and $\stackrel{-}{W}, \stackrel{-}{W'},
\stackrel{-}{W''}, \stackrel{-}{W'''}$ are four independent pairs
 of weight functions. It reduces to Baxter-Bazhanov model when
 $W=\stackrel{-}{W}$. The function $w(x,y,z|l)$ has the property
\begin{equation}
w(x,y,z|l)w(z,\omega^{1/2}y,\omega x|-l)\Phi(l)=1,\quad l\in Z_N.
\end{equation}
We will formulate the  weight functions of this model as the vertex
 form in the following section and find that the function
 $W(a|efg|bcd|h)$ is dependent on four spin variables which are
 the linear combinations of the spins $a,b,\cdots,h$.

\section{\bf Three-Dimensional Vertex Model}

In this section the three-dimensional vertex model is obtained.
 It is a duality of the 3D integrable lattice model presented above
 and  can be regarded as  a deformation of the vertex model given in
 ref. \cite{hujsp}.

Set
\begin{equation}
k_1=f+h-b-d,~~k_2=b+c-g-h,~~k_3=e+h-c-d,~~k_4=a+h-d-g.
\end{equation}
By taking account of the relation (5), we can express the weight
 function (1) as:
 $$
W(a|efg|bcd|h)=(-)^{k_2}(\omega^{1/2})^{k_1k_2+k_2k_3+k_1k_3}
\Bigg[\frac{w(x_1x_7,\omega^{1/2}x_3\xi,\omega x_3x_6|k_1)}
{w(x_4x_5,\omega^{1/2}x_4\lambda,\omega x_2x_8|k_4-k_3)}    ~~~
{}~~~~~~~
$$$$
{}~~~~~\times\frac{w(x_{58}x_{67},x_8\mu,x_{13}x_{24}x_7x_8/x_3x_4
|k_4-k_1-k_2-k_3)w(x_1x_8,x_1u,x_3x_5|k_3)}
{w(x_{58}x_{67},x_5\nu,x_{13}x_{24}x_5x_6/x_1x_2|-k_2)
w(x_4x_6,x_2\nu,x_2x_7|k_4-k_1)}\Bigg]^{1/2}
$$
\begin{equation} \label{W3}
\times \Bigg\{\sum_{\sigma \in Z_N}\frac{\omega^
{\sigma k_2}w(x_3,x_{13},x_1|\sigma)w(x_4,x_{24},x_2|k_4+\sigma)}
{w(x_8,x_{58},x_5|k_3+\sigma)w(x_7/\omega,x_{67},x_6|k_1+\sigma)
}\Bigg\}_0.
\end{equation}
It is obvious that the weight function $W(a|efg|bcd|h)$ depends
 on the spin variables $k_1,k_2,k_3,k_4$ defined in relation (6).
 Using the relation
$$
\Bigg\{\sum_{\sigma\in Z_N}\frac{w(x_3,x_{13},x_1|\sigma)
w(x_4,x_{24},x_2|\sigma+k_4)\omega^{\sigma k_2}}{w(x_8,x_{58},x_5
|\sigma +k_3)w(x_7/\omega,x_{67},x_6|\sigma+k_1)}\Bigg\}_0 ~~~~~~~
{}~~~~~~~~~~~~~~~~~~~~~~~~~~
$$$$
=\frac{w(x_4x_6,x_2v,x_2x_7|k_4-k_1)}{w(x_1x_8,x_1u,x_3x_5|k_3)} ~~~
{}~~~~~~~~~~~~~~~~~~~~~~~~~~~~~~~~~~~~~~~~~~~~~~~~
$$
\begin{equation}
{}~~~~~~~~~~~~~\times \Bigg\{\sum_{\sigma\in Z_N}
\frac{w(x_8x_{13},x_1u,x_3x_{58}|\sigma+k_3+k_2)
w(x_2x_{67},x_2v,x_6x_{24}|\sigma)\omega^
{\sigma k_1}}{w(x_5x_{13},\omega x_1u,\omega x_1x_{58}|\sigma+k_2)
w(x_4x_{67},x_2v,x_7x_{24}|\sigma+k_4-k_1)}\Bigg\}_0
\end{equation}
and fixing the normalizations of all parameters $x's$ appeared in
 the expression (1) as (see ref.\cite{boose})
\begin{equation}
x_3=x_4=x_7=x_8=1,
\end{equation}
we can change the weight function (7) into the vertex form
$$
R^{j_1j_2j_3}_{i_1i_2i_3}=(-)^{j_2}(\omega^{1/2})^
{j_1j_2+j_2j_3+j_1j_3} ~~~~~~~~~~~~~~~~~~~~~~~~~~~~~~~~~~~~~
{}~~~~~~~~~~~~~~~~~~~~~~
$$
$$
{}~~~\times\Bigg[\frac{w(x_1,\omega^{1/2}\xi,\omega x_6|j_1)
w(x_{58}x_{67},\mu,x_{13}x_{24}|-i_2)w(x_6,x_2v,x_2|i_3)}
{w(x_5,\omega^{1/2}\lambda,\omega x_2|i_1)
w(x_{58}x_{67},x_5\nu,x_{13}x_{24}x_5x_6/x_1x_2|-j_2)
w(x_1,x_1u,x_5|j_3)}\Bigg]^{1/2}
$$
\begin{equation} \label{R2}
{}~~~~~~~~~~~~~\times \Bigg\{\sum_{\sigma\in Z_N}
\frac{w(x_{13},x_1u,x_{58}|\sigma+j_2+j_3)
w(x_2x_{67},x_2v,x_6x_{24}|\sigma)s(\sigma,j_1)}
{w(x_5x_{13},\omega x_1u,\omega x_1x_{58}|\sigma+j_2)
w(x_{67},x_2v,x_{24}|\sigma+i_3)}\Bigg\}_0 ~~
\end{equation}
where
\begin{equation} \begin{array}{ll}
i_1=k_4-k_3, & j_1=k_1,\\[0.3cm]
i_2=k_1+k_2+k_3-k_4, & j_2=k_2,\\[0.3cm]
i_3=k_4-k_1, & j_3=k_3.
\end{array}
\end{equation}
Set
\begin{equation}
\frac{x_5}{\omega x_2}=q^{-1}u_1,~~\frac{u'_2}{u''_2}=qu_2,
{}~~\frac{x_2x_{13}}{\omega x_6x_{58}}=qu'_2,~~\frac{x_2x_{67}}
{x_6x_{24}}=qu_2'',~~\frac{x_6}{x_2}=q^{-1}u_3
\end{equation}
where
\begin{equation}
x_1x_2=q^2x_5x_6.
\end{equation}
By considering the notation \cite{BB1}
\begin{equation}
{w(v,a)\over w(v,0)}=[\Delta(v)]^a\prod^a_{j=1}(1-\omega^jv)^{-1},
{}~~\Delta(v)=(1-v^N)^{1/N},
\end{equation}
the weight function has the form
$$
R(u_1,u_2,u_3)^{j_1j_2j_3}_{i_1i_2i_3}=(-)^{j_2}(\omega^{1/2})^
{j_1j_2+j_2j_3+j_1j_3}~~~~~~~~~~~~~~~~~~~~~~~~~~~~~~~~~~~~
$$$$
{}~~\times\Bigg[\frac{w(qu_1,j_1)w(q^{-1}(\omega u_2)^{-1},-i_2)
w(q^{-1}u_3,i_3)}{w(q^{-1}u_1,i_1)w(q(\omega u_2)^{-1},-j_2)
w(qu_3,j_3)}\Bigg]^{1/2}
$$
\begin{equation}
{}~~~~~\times\Bigg\{\sum_{\sigma\in Z_N}
\frac{w(\omega u'_2u_3,\sigma+j_2+j_3)w(qu_2'',\sigma)
s(\sigma,j_1)}{w(q^{-1}u'_2,\sigma+j_2)w(u_2''u_3,\sigma+i_3)}
\Bigg\}_0.
\end{equation}
Similarly as in ref. \cite{boose}, we can express the spectrums
 $u_i~~(i=1,2,3)$  as
\begin{equation}
 u^N_i=\frac{C_i+D_i}{C_i-D_i},
\end{equation}
where $C_i,~D_i$ satisfy the relation
\begin{equation}
\frac{1-D^2_i}{1-C^2_i}=\frac{(1-q^N)^2}{(1+q^N)^2},~~i=1,2,3.
\end{equation}
Then let
\begin{equation}
cos(a_0)=\frac{k^2C_1C_2C_3-D_1D_2D_3}{k'^2},~~cos(a_r)=
\frac{(-)^r(D_pD_qC_r-C_pC_qD_r)}{k'^2S_pS_q},
\end{equation}
where $\{p,q,r\}=\{1,2,3\}$ with
 \begin{equation}
C_i^2+S_i^2=1,~~k^2+k'^2=1,~~k=\frac{1-q^N}{1+q^N}.
\end{equation}
We can denote the variables $x_i^N$ as
\begin{equation}\begin{array}{l}
x^N_1=-(u_1u_3)^{N/2}\sqrt{\frac{1-(qu_2)^N}{1-(q^{-1}u_2)^N}}
 exp(ia_2),\\[0.4cm]
x^N_2=-(u_1u_3)^{-N/2}\sqrt{\frac{1-(qu_2)^N}{1-(q^{-1}u_2)^N}}
 exp(ia_2),\\[0.4cm]
x^N_5=-q^{-N}(u_1u_3^{-1})^{N/2}\sqrt{\frac{1-(qu_2)^N}{1-(q^{-1}u_2)^N}}
 exp(ia_2),\\[0.4cm]
x^N_6=-q^{-N}(u_1^{-1}u_3)^{N/2}\sqrt{\frac{1-(qu_2)^N}{1-(q^{-1}u_2)^N}}
 exp(ia_2),
\end{array}\end{equation}
where the phases are chosen by
\begin{equation}
\frac{x_1}{|x_1|}=\frac{x_2}{|x_2|}=\frac{x_5}{\omega^{1/2}|x_5|}=
\frac{\omega^{1/2}x_6}{|x_6|}=e^{ia_2/N}.
\end{equation}
And the expressions of the variables $x_{13},x_{24},x_{58},x_{67}$
 can be obtained by
$$
x_{13}^N=x_1^N-1, ~~x_{24}^N=x_2^N-1, ~~x_{58}^N=x_5^N-1,~~x_{67}^N=x_6^N-1
$$
from relations (20) with the phases
\begin{equation}\begin{array}{ll}
\frac{\displaystyle{x_{13}}}{\displaystyle{|x_{13}|}}=
e^{i(a_0+a_1+a_2+a_3)/(2N)}, &
\frac{\displaystyle{x_{24}}}{\displaystyle{|x_{24}|}}=
\omega^{1/2}e^{i(a_0-a_1+a_2-a_3)/(2N)},\\[0.4cm]
\frac{\displaystyle{x_{58}}}{\displaystyle{|x_{58}|}}=
\omega^{1/2}e^{i(a_0-a_1+a_2+a_3)/(2N)}, &
\frac{\displaystyle{x_{67}}}{\displaystyle{|x_{67}|}}=
\omega^{-1/2}e^{i(a_0+a_1+a_2-a_3)/(2N)}.
\end{array}\end{equation}
In this way, the weight function (15) satisfies the modified vertex
 type tetrahedron equation
$$
{\displaystyle\sum_{\{k_i\},\atop i=1,\cdots,6}}
R(u_1,u_2,u_3)^{k_1,k_2,k_3}_{i_1,i_2,i_3}\overline{R}(u_1,u_4,u_5)^
{j_1k_4k_5}_{k_1i_4i_5}
R(u_2,u_4,u_6)^{j_2j_4k_6}_{k_2k_4i_6}
\overline{R}(u_3,u_5,u_6)^{j_3j_5j_6}_{k_3k_5k_6}= ~~~~~~~
$$
\begin{equation}
{}~{\displaystyle\sum_{\{k_i\},\atop i=1,\cdots,6}}
R(u_3,u_5,u_6)^{k_3,k_5,k_6}_{i_3,i_5,i_6}
\overline{R}(u_2,u_4,u_6)^{k_2k_4j_6}_{i_2i_4k_6}
R(u_1,u_4,u_5)^{k_1j_4j_5}_{i_1k_4k_5}
\overline{R}(u_1,u_2,u_3)^{j_1j_2j_3}_{k_1k_2k_3}
\end{equation}
where $\overline{R}(u_1,u_2,u_3)$ can be obtained from $R(u_1,u_2,u_3)$
by the substitutions:
\begin{equation}
q\rightarrow q^{-1},~~u_2'\rightarrow \overline{u}_2', ~~u_2''\rightarrow
 \overline{u}_2'',
\end{equation}
and
\begin{equation}\begin{array}{l}
u_1=\frac{\displaystyle{qx_5}}{\displaystyle{\omega x_2}}=
\frac{\displaystyle{q^{-1}\overline{x}_5}}
{\displaystyle{\omega\overline{x}_2}}
=\frac{\displaystyle{qx_5'}}{\displaystyle{\omega x_2'}}=
\frac{\displaystyle{q^{-1}\overline{x}_5'}}
{\displaystyle{\omega\overline{x}_2'}},\\[0.4cm]
u_2=\frac{\displaystyle{q^{-1}x_{13}x_{24}}}
{\displaystyle{\omega x_{58}x_{67}}}=
\frac{\displaystyle{q\overline{x}_{13}\overline{x}_{24}}}
{\displaystyle{\omega\overline{x}_{58}\overline{x}_{67}}}
=\frac{\displaystyle{qx_5''}}{\displaystyle{\omega x_2''}}=
\frac{\displaystyle{q^{-1}\overline{x}_5''}}
{\displaystyle{\omega\overline{x}_2''}},\\[0.4cm]
u_3=\frac{\displaystyle{qx_6}}{\displaystyle{x_2}}=
\frac{\displaystyle{q^{-1}\overline{x}_6}}
{\displaystyle{\overline{x}_2}}
=\frac{\displaystyle{qx_5'''}}{\displaystyle{\omega x_2'''}}=
\frac{\displaystyle{q^{-1}\overline{x}_5'''}}
{\displaystyle{\omega\overline{x}_2'''}},\\[0.4cm]
u_4=\frac{\displaystyle{q^{-1}x_{13}'x_{24}'}}
{\displaystyle{\omega x_{58}'x_{67}'}}=
\frac{\displaystyle{q\overline{x}_{13}'\overline{x}_{24}'}}
{\displaystyle{\omega\overline{x}_{58}'\overline{x}_{67}'}}
=\frac{\displaystyle{q^{-1}x_{13}''x_{24}''}}
{\displaystyle{\omega x_{58}''x_{67}''}}=
\frac{\displaystyle{q\overline{x}_{13}''\overline{x}_{24}''}}
{\displaystyle{\omega\overline{x}_{58}''\overline{x}_{67}''}},\\[0.4cm]
u_5=\frac{\displaystyle{qx_6'}}{\displaystyle{x_2'}}=
\frac{\displaystyle{q^{-1}\overline{x}_6'}}
{\displaystyle{\overline{x}_2'}}
=\frac{\displaystyle{q^{-1}x_{13}'''x_{24}'''}}
{\displaystyle{\omega x_{58}'''x_{67}'''}}=
\frac{\displaystyle{q\overline{x}_{13}'''\overline{x}_{24}'''}}
{\displaystyle{\omega\overline{x}_{58}'''\overline{x}_{67}'''}},\\[0.4cm]
u_6=\frac{\displaystyle{qx_6''}}{\displaystyle{x_2''}}=
\frac{\displaystyle{q^{-1}\overline{x}_6''}}
{\displaystyle{\overline{x}_2''}}
=\frac{\displaystyle{qx_6'''}}{\displaystyle{x_2'''}}=
\frac{\displaystyle{q^{-1}\overline{x}_6'''}}
{\displaystyle{\overline{x}_2'''}}.
\end{array}\end{equation}
Notice that the spectrums should satisfy the conditions:
\begin{equation}\begin{array}{c}
a_0+a_1'-a_1''+a_1'''=a_1+a_0'-a_2''+a_2'''=0,\\[0.4cm]
a_2-a_2'+a_0''+a_3'''=a_3-a_3'+a_3''+a_0'''=0,
\end{array}\end{equation}
where $a_i',a_i'',a_i''' ~(i=0,1,2,3)$ are defined similarly as in
relations (16) and (18) (see ref.\cite{boose}). So we get a
three-dimensional vertex model which is a duality of the BCC model.
It is a deformation of the three-dimensional vertex model in ref.
 \cite{hujsp}. The details will be given in the following section.
 And the constrained conditions of the spectrums between them are
 discussed.

\section {The Additional Constraints of the Parameters}

By setting q=1, the weight function (15) becomes as
$$
R(u_1,u_2,u_3)^{j_1j_2j_3}_{i_1i_2i_3}=(-)^{j_2}(\omega^{1/2})^
{j_1j_2+j_2j_3+j_1j_3}~~~~~~~~~~~~~~~~~~~~~~~~~~~~~~~~~~~~
$$$$
{}~~\times\Bigg[\frac{w(u_1,j_1)w((\omega u_2)^{-1},-i_2)w(u_3,i_3)}
{w(u_1,i_1)w((\omega u_2)^{-1},-j_2)w(u_3,j_3)}\Bigg]^{1/2}
$$
\begin{equation}
{}~~~~~\times\Bigg\{\sum_{\sigma\in Z_N}
\frac{w(\omega u'_2u_3,\sigma+j_2+j_3)w(u_2'',\sigma)
s(\sigma,j_1)}{w(u'_2,\sigma+j_2)w(u_2''u_3,\sigma+i_3)}\Bigg\}_0.
\end{equation}
It is justly the weight function of the 3D vertex  model \cite{hujsp}
related Baxter-Bazhanov model. In this case the modified vertex type
 tetrahedron equation  (23) reduces to the ordinary one. We know that
 the four additional constraints are
\begin{equation}\begin{array}{rr}
\omega\frac{\displaystyle{x_{23}}}{\displaystyle{x_3}}
\frac{\displaystyle{x_4'}}{\displaystyle{x_{24}'}}
\frac{\displaystyle{x_{24}''}}{\displaystyle{x_2''}}
\frac{\displaystyle{x_2'''}}{\displaystyle{x_{24}'''}}=1,&
\frac{\displaystyle{x_{13}}}{\displaystyle{x_1}}
\frac{\displaystyle{x_1'}}{\displaystyle{x_{14}'}}
\frac{\displaystyle{x_{14}''}}{\displaystyle{x_1''}}
\frac{\displaystyle{x_1'''}}{\displaystyle{x_{14}'''}}=1,\\[6mm]
\frac{\displaystyle{x_{14}}}{\displaystyle{x_4}}
\frac{\displaystyle{x_4'}}{\displaystyle{x_{14}'}}
\frac{\displaystyle{x_{14}''}}{\displaystyle{x_4''}}
\frac{\displaystyle{x_4'''}}{\displaystyle{x_{24}'''}}=1,&
\frac{\displaystyle{x_{13}}}{\displaystyle{x^{\*}_3}}
\frac{\displaystyle{x_3'}}{\displaystyle{x_{13}'}}
\frac{\displaystyle{x_{13}''}}{\displaystyle{x_1''}}
\frac{\displaystyle{x_2'''}}{\displaystyle{x_{23}'''}}=1,
\end{array}\end{equation}
for the case of $q=1$. By making the transformations:
\begin{equation}
\frac{x_1}{x_2}\rightarrow \frac{x_1}{x_6},~~
\frac{x_1}{x_3}\rightarrow x_1,\frac{x_4}{x_3}\rightarrow
\frac{x_1}{x_5},
\end{equation}
and choosing that
\begin{equation}
\frac{x_{23}}{x_{13}}\rightarrow\frac{x_{67}}{x_{13}},
{}~~\frac{x_{13}}{x_1}\rightarrow\frac{x_{13}}{x_1}, ~~
\frac{x_{13}}{x_{14}}\rightarrow\frac{x_5x_{13}}{x_1x_{58}},~~
\frac{x_{24}}{x_{23}}\rightarrow\frac{x_1x_{24}}{x_5x_{67}},
\end{equation}
we have
$$
\frac{x_{13}}{x_3}\rightarrow x_{13},~~
\frac{x_{14}}{x_1}\rightarrow\frac{x_{58}}{x_5},~~
\frac{x_{23}}{x_2}\rightarrow\frac{x_{67}}{x_6},~~
\frac{x_{23}}{x_3}\rightarrow x_{67},
$$
\begin{equation}
\frac{x_{14}}{x_4}\rightarrow x_{58},~~
\frac{x_{24}}{x_2}\rightarrow\frac{x_{24}}{x_2},~~
\frac{x_{24}}{x_4}\rightarrow x_{24}.
\end{equation}
Using the transformations (29), (30) and (31), we can change (28)
into the form:
\begin{equation}\begin{array}{rr}
\omega\frac{\displaystyle{x_{67}}}{\displaystyle{x_{24}'}}
\frac{\displaystyle{x_{24}''}}{\displaystyle{x_2''}}
\frac{\displaystyle{x_2'''}}{\displaystyle{x_{24}'''}}=1,&
\frac{\displaystyle{x_{13}}}{\displaystyle{x_1}}
\frac{\displaystyle{x_5'}}{\displaystyle{x_{58}'}}
\frac{\displaystyle{x_{58}''}}{\displaystyle{x_5''}}
\frac{\displaystyle{x_5'''}}{\displaystyle{x_{58}'''}}=1,\\[4mm]
\frac{\displaystyle{x_{58}}}{\displaystyle{x_{58}'}}
\frac{\displaystyle{x_{58}''}}{\displaystyle{x_{24}'''}}=1,&
\frac{\displaystyle{x_{13}}}{\displaystyle{x_{13}'}}
\frac{\displaystyle{x_{13}''}}{\displaystyle{x_1''}}
\frac{\displaystyle{x_6'''}}{\displaystyle{x_{67}'''}}=1.
\end{array}\end{equation}
By comparing with the constraints for the modified tetrahedron
 equation introduced  in ref. \cite{boose},
\begin{equation}\begin{array}{llc}
\omega\frac{\displaystyle{x_{67}}}{\displaystyle{\overline{x}_{24}'}}
\frac{\displaystyle{x_{24}''}}{\displaystyle{x_2''}}
\frac{\displaystyle{x_2'''}}{\displaystyle{x_{24}'''}}=1,&
\frac{\displaystyle{\overline{x}_{13}}}{\displaystyle{\overline{x}_1}}
\frac{\displaystyle{\overline{x}_1'}}{\displaystyle{\overline{x}_{13}'}}
\frac{\displaystyle{\overline{x}_{13}''}}
{\displaystyle{\overline{x}_1''}}\frac{\displaystyle{\overline{x}_1'''}}
{\displaystyle{\overline{x}_{13}'''}}=1,&\\[4mm]
\frac{\displaystyle{x_{58}}}{\displaystyle{x_{58}'}}
\frac{\displaystyle{x_{58}''}}{\displaystyle{\overline{x}_{24}'''}}=1,&
\omega\frac{\displaystyle{x_{67}'}}{\displaystyle{\overline{x}_{24}}}
\frac{\displaystyle{x_{67}'''}}{\displaystyle{x_{67}''}}=1, &
\frac{\displaystyle{x_{24}}}{\displaystyle{x_{24}'}}
\frac{\displaystyle{x_{24}''}}{\displaystyle{x_{24}'''}}=1,
\end{array}\end{equation}
we get that the two relations in the left hand sides of (32) are
corresponding to the left ones in (33) with $q=1$. It can be checked
easily that the two relations in the right hand sides of (32) hold
also by using relations (20), (21) and (22) for the case of $q=1$.
 And the other three relations in (33) become as
\begin{equation}
\frac{\displaystyle{x_{13}}}{\displaystyle{x_1}}
\frac{\displaystyle{x_1'}}{\displaystyle{x_{13}'}}
\frac{\displaystyle{x_{13}''}}{\displaystyle{x_1''}}
\frac{\displaystyle{x_1'''}}{\displaystyle{x_{13}'''}}=1, ~~
\omega\frac{\displaystyle{x_4}}{\displaystyle{x_{24}}}
\frac{\displaystyle{x_{23}'}}{\displaystyle{x_3'}}
\frac{\displaystyle{x_3''}}{\displaystyle{x_{23}''}}
\frac{\displaystyle{x_{23}'''}}{\displaystyle{x_3'''}}=1, ~~
\frac{\displaystyle{x_{24}}}{\displaystyle{x_4}}
\frac{\displaystyle{x_4'}}{\displaystyle{x_{24}'}}
\frac{\displaystyle{x_{24}''}}{\displaystyle{x_4''}}
\frac{\displaystyle{x_4'''}}{\displaystyle{x_{24}'''}}=1,
\end{equation}
by considering the variables substitutions (29), (30) and (31). These
relations are correct indeed for Baxter-Bazhanov model and they are
 equivalent to the relation (36) of ref. \cite{hujsp}. The spectrums
 in the modified vertex type tetrahedron equation (23) can be
 parameterized as
\begin{equation}
u_i=\omega^{-1/2}[tn(\theta_i/2,k)]^{-2/N},~~~~~~i=1,2,\cdots,6,
\end{equation}
where
\begin{equation}
tn(\frac{\theta_i}{2},k)= \frac{k' sn(\theta_i/2,k)}{cn(\theta_i/2,k)
dn(\theta_i/2,k)}.
\end{equation}
When $q=1$, that is, $k=0$, they become as
\begin{equation}
u_i=\omega^{-1/2}[ctg(\frac{\theta_i}{2})]^{2/N}, ~~~~~~i=1,2,\cdots,6.
\end{equation}
And the model reduces to the 3D vertex model presented in ref.
\cite{hujsp}.

\section{Symmetry Properties and Some Remarks}

Let us denote the two generating elements of the group $G$ consisting
 of various rotations, reflections and their combinations of the cube
as $\tau$ and $\rho$ \cite{mangazeev,hujsp}. Under these
 transformations the parameters of the weight function change
 as (see ref.\cite{boose})
\begin{equation}
\frac{x_5}{\omega x_2}\stackrel{\tau}{\longleftrightarrow}
\frac{x_6}{x_2},~~
\frac{x_{13}x_{24}}{x_{58}x_{67}}\stackrel{\tau}{\longleftrightarrow}
\frac{x_{13}x_{24}}{x_{58}x_{67}}
\end{equation}
and
$$
\frac{x_1}{x_5}\stackrel{\rho}{\longleftrightarrow}\frac{x_5}
{\omega x_1},~~\frac{x_2}{x_6}\stackrel{\rho}{\longleftrightarrow}
\frac{\omega x_6}{x_2},~~\frac{x_1}{x_6}\stackrel{\rho}
{\longleftrightarrow}\frac{\omega x_{58}x_{67}}{x_{13}x_{24}},
$$
\begin{equation}
\frac{x_{13}}{x_{58}}\stackrel{\rho}{\longleftrightarrow}\frac{x_5}
{\omega},~~~~
\frac{x_{24}}{x_{67}}\stackrel{\rho}{\longleftrightarrow}
\frac{\omega}{x_2}.
\end{equation}
Then the symmetry properties of the weight function (15) are
\begin{equation}
R(u_1,u_2,u_3)^{j_1j_2j_3}_{i_1i_2i_3}\stackrel{\tau}{=}
R(u_3,u_2,u_1)^{j_3j_2j_1}_{i_3i_2i_1},
\end{equation}
\begin{equation}
R(u_1,u_2,u_3)^{j_1j_2j_3}_{i_1i_2i_3}\stackrel{\rho}{=}
R((\omega u_2)^{-1},u_1,(\omega u_3)^{-1})^{-i_2j_1-j_3}_{-j_2i_1-i_3}.
\end{equation}
When $q=1$, the above two relations reduce to the relations (44)
 and (45) in ref. \cite{hujsp}. From these properties we have
\begin{equation}
R(u_1,u_2,u_3)^{j_1j_2j_3}_{i_1i_2i_3}=R((\omega u_3)^{-1},
(\omega u_2)^{-1},u_1)^{-j_3-j_2i_1}_{-i_3-i_2j_1}.
\end{equation}
It is a deformation of the star-star relation of the Baxter-Bazhanov
 model and reduces to the relation (39) of ref. \cite{hujsp} when $q=1$.

As the conclusions, we obtained the three-dimensional vertex model.
The weight function of the model depends on four spins variables
 which are the linear combinations of the spins on the corner sites
 of the cube. The three spectrums of the vertex type weight function
 (15) relate the three `space'  of the function $R$ where a
deformation parameter $q$ exists. The weight functions satisfy the
modified vertex type tetrahedron equation where the six parameters
 are corresponding to six space on which the tetrahedron equation is
 defined. This vertex model is a duality of the BCC model. When $q=1$,
 the vertex model reduces to the one introduced in ref. \cite{hujsp}.
 The constraints are discussed also in detail. The symmetry properties
 of the weight function are given, from which we can get all of the
 symmetry transformations of the vertex type weight function under the
 action of the group $G$ consisting of the symmetrical maps of the
three-dimensional cube. One of them is the deformation of the
three-dimensional star-star relation of the Baxter-Bazhanov model.

Very recently, the modified tetrahedron equation was discussed also
in ref.\cite{boosnew} and the vertex formulation of the Bazhanov-Baxter
 model is given ,too, in ref. \cite{sms}. We know that this 3D lattice
 model was constructed  firstly through the study of the chiral Potts
 model. And the chiral Potts model can be regarded as a descendant
 of the six-vertex model in two dimensions \cite{lieb,bsjsp}. The
hidden symmetries of the six-vertex model  transfer matrix and the
 correlation function \cite{jimbo,z3} of the $XXZ$ chain was
 discussed  in ref.\cite{kuan,pin} recently. What is the case in
 three dimensions? It is a interesting problem to discuss the
correlation function and the free energy \cite{phyD} for the
 three-dimensional integrable model. I hope the above discussions
 will be useful also for finding an algebra defined by
 $RLLL=LLLR$ similarly as the fusion of the Yang-Baxter equation
 in two dimensions.

\section*{\bf Acknowledgment}

The author would like to thank B. Y. Hou, K. J. Shi, Z. B. Su,
 P. Wang
and K. Wu for the interesting discussions and J. Hietarinta, I.
 Korepanov and V.  Korepin for the interests of this work. I am also
grateful B. L. Hao  for his reports at the meeting of statistical
 physics and nonlinear phenomena.

\end{document}